\documentclass[aps,preprint,floatfix,nofootinbib,showpacs,superscriptaddress]{revtex4-1}
\pdfoutput=1
 
\catcode`\@=11
\def\lsim{\mathrel{\mathpalette\@versim<}}
\def\gsim{\mathrel{\mathpalette\@versim>}}
\def\@versim#1#2{\vcenter{\offinterlineskip
		\ialign{$\m@th#1\hfil##\hfil$\crcr#2\crcr\sim\crcr } }}
\catcode`\@=12

\catcode`@=12
\usepackage[export]{adjustbox}	
\usepackage{float}
\usepackage{epsfig,bbm}
\usepackage{amsmath}
\usepackage{slashed}
\usepackage{tikz}
\usepackage{bm}
\usepackage{amssymb}
\usepackage{mathtools}
\usepackage[caption=false]{subfig}
\usepackage{graphicx}
\usepackage{hyperref}
\usepackage{cleveref}
\usepackage{tabularx}

\usepackage[normalem]{ulem}

\usepackage{lineno}

\def\lsim{\mathrel{\rlap{\lower4pt\hbox{\hskip1pt$\sim$}}
		\raise1pt\hbox{$<$}}} 
\def\gsim{\mathrel{\rlap{\lower4pt\hbox{\hskip1pt$\sim$}}
		\raise1pt\hbox{$>$}}} 

\newcommand{\newc}{\newcommand}
\newc{\renewc}{\renewcommand}
\newc{\red}{\textcolor{red}}
\newc{\blue}{\textcolor{blue}}
\newc{\psh}{\textcolor{psh}}
\newc{\yjk}{\textcolor{dG}}
\newcolumntype{C}[1]{>{\centering\arraybackslash}p{#1}}

\def\JPsi{J/\Psi}
\def\epem{e^+ e^-}
\def\lplm{\ell^+ \ell^-}

\def\VtoPepem{V \to P \epem}
\def\etacepem{\eta_c \epem}

\definecolor{dG}{rgb}{0.1, 0.5, 0}
\definecolor{Navy}{rgb}{0.0, 0.2, 0.6}
\definecolor{sRed}{rgb}{0.6,0.1,0.1} 
\definecolor{psh}{rgb}{0.5, 0, 0.3} 

\def\Mreceell{M^{\mathrm{rec}}_{ee\ell\ell}}

\begin{document}
	\thispagestyle{empty}

	\title{\Large Search for new light vector boson \\
	using $J/\Psi$	 at BESIII and Belle~II}

	\author{Kayoung Ban}
	\email{ban94gy@yonsei.ac.kr}
	\affiliation{Department of Physics and IPAP, Yonsei University, \\
		Seoul 03722, Republic of Korea}

	\author{Yongsoo Jho}
	\email{jys34@yonsei.ac.kr}
	\affiliation{Department of Physics and IPAP, Yonsei University, \\
		Seoul 03722, Republic of Korea}

	\author{Youngjoon Kwon}
	\email{yjkwon63@yonsei.ac.kr }
	\affiliation{Department of Physics and IPAP, Yonsei University, \\
		Seoul 03722, Republic of Korea}
	
	\author{Seong Chan Park}
	\email{sc.park@yonsei.ac.kr}
	\affiliation{Department of Physics and IPAP, Yonsei University, \\
		Seoul 03722, Republic of Korea}
	
	\author{Seokhee Park}
	\email{seokhee.park@yonsei.ac.kr}
	\affiliation{Department of Physics and IPAP, Yonsei University, \\
		Seoul 03722, Republic of Korea}

	\author{Po-Yan Tseng}
	\email{tpoyan1209@gmail.com}
	
	\affiliation{Department of Physics and IPAP, Yonsei University, \\
		Seoul 03722, Republic of Korea}
	

	\preprint{YHEP-COS20-06  
	}

	\begin{abstract}
		
		We investigate various search strategies for light vector boson $X$ in $\mathcal{O}(10)~{\rm MeV}$ mass range using $J/\Psi$ 
		associated channels at BESIII and Belle~II: 
		{\it (i)} $J/\Psi \to \eta_c X$ with $10^{10} J/\Psi$s at BESIII, {\it (ii)} $J/\Psi (\eta_c +X) +\ell \bar{\ell}$ production at Belle~II,  and {\it (iii)} $J/\Psi +X$  with the displaced vertex in $X\to e^+e^-$ decay are analyzed and the future sensitivities at Belle~II with 50~${\rm ab}^{-1}$ luminosity are comprehensively studied. 
		By requiring the displaced vertex to be within the beam pipe, the third method results in nearly background-free analysis, and the vector boson-electron coupling and the vector boson mass can be probed in the unprecedented range,  
		$10^{-4}\leq |\varepsilon_e| \leq  10^{-3}$
		and $9~{\rm MeV}\leq m_X\leq 100 {\rm MeV}$
		with 50 ${\rm ab}^{-1}$ at Belle~II.  
		This covers the favored signal region of $^8{\rm Be}^*$ anomaly recently reported by Atomki experiment with $m_X \simeq 17~{\rm MeV}$.

	\end{abstract}
	

	\maketitle

	\section{Introduction}
	
  The Standard Model (SM) is a successful theory describing physics at least up to the electroweak scale, having survived more than 40 years of various experimental tests.
{  However, there still remain a handful number of experimental and observational claims that indicate discrepancies from the SM predictions and consequently request extension of the SM: non-zero mass of neutrinos~\cite{Fukuda:1998mi},  anomalous magnetic moment of muon, $(g-2)_\mu$~\cite{Bennett:2006fi, Keshavarzi:2019abf},
existence of dark matter (DM)~\cite{Aghanim:2018eyx},~\footnote{Primordial black holes may explain the whole DM. See, e.g., \cite{Cheong:2019vzl}.} and baryon vs. antibaryon asymmetry of the universe \cite{Canetti:2012zc, Zyla:2020zbs}.}
	There have been discussions of extending the
	SM by gauging the lepton number, 
	e.g. $L_\mu-L_\tau$ or $L_e-L_\tau$~\cite{He:1991qd}, 
	 intending to explain DM~\cite{Altmannshofer:2016jzy, Foldenauer:2018zrz}, the muon anomalous magnetic moment $(g-2)_{\mu}$~\cite{Baek:2001kca, Kaneta:2016uyt, Araki:2017wyg, Jho:2019cxq}, and {more recently} EDGES 21cm anomaly~\cite{Berlin:2018sjs}. 	
	The extension gives rise to a leptophilic light vector boson, dubbed as $X$ in this paper.
	{We note that the $X$ boson may couple to the quarks via interactions 
	with unknown heavy fermions that mix with SM quarks~\cite{Altmannshofer:2014cfa}.
  It may then be responsible for the recent anomaly from the KOTO experiment 
	in $K_L \to \pi^0 \nu \bar{\nu}$~\cite{Shinohara:2019, Tung:2019, Lin:2019,Jho:2020jsa} and also 
  the anomaly from the Atomki experiment in both $^8{\rm Be}^*$ and $^4{\rm He}^*$. The preferred mass of $X$ for these cases is in sub-GeV range; in particular $m_X\simeq 17~{\rm MeV}$ for Atomki~\cite{Krasznahorkay:2015iga,Krasznahorkay:2019lyl}.

	{High luminosity lepton colliders provide ideal environments to test for such light $X$ boson. 
  Thanks to less severe QCD backgrounds, the lepton colliders have definite advantages over hadron colliders 
	even when the $X$ boson has feeble couplings with the SM particles.
	 In this paper, we take the lepton colliders, BESIII and Belle~II, and study the search strategies of $X$. 
   In particular, we focus on the channels in association with a $J/\Psi$ meson, which 
	 will be enormously produced at BESIII and also at Belle~II, thereby leaving the signals of $X$ in various channels: 
	
	\begin{itemize}
\item	 At BESIII, up to now,  $10^{10}$ $J/\Psi$ events are collected, thus provide an excellent probe to study the $J/\Psi$ rare decays to the $X$ boson.
	
\item	At Belle~II, even though less number of $J/\Psi$ are expected, we use the process 
	$e^+e^- \to \ell^+\ell^- J/\Psi \to \ell^+\ell^- \eta_c X \to \ell^+\ell^- \eta_c e^+ e^- $ {($\ell = e$ or $\mu$)}, in which $J/{\Psi}$ and $\eta_c$ are inferred by the recoil masses of $\ell^+\ell^-$ and $\ell^+ \ell^- e^+ e^-$, respectively.
	
\item At Belle~II, we also use the channel $e^+e^- \to X+J/\Psi$ where the $X$ bosons will leave signals with displaced vertices. 
        Due to higher center-of-mass (CM) energy at Belle~II, the $X$ boson will be boosted and travel several millimeters before it decays into $e^+e^-$. 
        
       	\end{itemize}
  The rest of this paper is dedicated to studying the sensitivity reach of finding $X$ boson taking realistic experimental situations into account
  under the effective field theory framework.

	This paper is organized as follows: we first set up our theoretical framework and introduce the effective interactions in Section~\ref{sec:interaction}.
  The analysis for BESIII is carried out in Section~\ref{sec:spectrum}. 
	In Section~\ref{sec:4lepton} and ~\ref{sec:displaced}, $e^+e^- \to \ell^+\ell^- J/\Psi$ 
	and $e^+e^- \to X+J/\Psi$ with displaced vertex signal are analyzed, respectively for Belle~II. 
	Finally, our results are summarized in Section~\ref{sec:summary}.

	\section{Effective Lagrangian}
	\label{sec:interaction}

	{The vectorlike  interactions of the $X$ boson with the SM fermions, $f$, are introduced by 
	the effective Lagrangian:
	\begin{align}
	\mathcal{L} 
	&\supset -e X_\mu \sum_f \varepsilon_f \bar{f}\gamma^\mu f, 
	\end{align}
  where we regard the couplings $\varepsilon_f$ as free parameters without knowing the origin.  	In particular, we will assume four universal couplings, $\varepsilon_u, \varepsilon_d, \varepsilon_e$, and $\varepsilon_\nu$, for
  up-type quarks, down-type quarks, charged leptons, and neutrinos, respectively, in our analysis below.  We also note that the new interactions do not induce any axial anomaly by construction.}

		{If the new boson $X$ is responsible for the recent Atomki anomaly~\cite{Krasznahorkay:2019lyl} via
	the process ${\rm^8Be}+ X \to {\rm ^8Be}+ e^+e^-$, its mass should be $m_X\simeq 17$ MeV and the couplings with the first generation quarks should be in a particular window ~\cite{Feng:2016jff,Fornal:2017msy,Feng:2020mbt}:
	\begin{align}
	\label{eq:parameter}
		|\varepsilon_u+\varepsilon_d| &\simeq  3.7\times 10^{-3}. 
	\end{align}
      From the NA48/2 experiment for $\pi^0 \to X \gamma$, we require a {\it protophobic} condition~\cite{Batley:2015lha}:
	\begin{eqnarray}
		\label{eq:parameter2}
	|2\varepsilon_u+\varepsilon_d|<8\times 10^{-4}.
	\end{eqnarray}
  Taking both relations in Eq.~\ref{eq:parameter} and Eq.~\ref{eq:parameter2} into account, we finally get the preferred value for  up-type and down-type quark couplings:
	\begin{align}
	\varepsilon_u \simeq  \pm 3.7\times 10^{-3},\,	\varepsilon_d \simeq  \mp 7.4\times 10^{-3},
	\end{align}
	which we will rely on below. 
	}

	{The coupling to the leptons, especially to electron and electron-neutrino, are stringently constrained by the beam dump experiment SLAC E141~\cite{Essig:2013lka},  the anomalous magnetic moment of the electron $g-2$~\cite{Davoudiasl:2014kua}, and neutrino-electron scattering experiment~\cite{Khan:2016uon}:  
	\begin{align}
	4.2\times 10^{-4} \lesssim|\varepsilon_e| & \lesssim 1.4\times 10^{-3}, \\
	\sqrt{\varepsilon_e \varepsilon_\nu}  &\lesssim 7 \times 10^{-5}.
	\end{align}
	When a small coupling for neutrino $\varepsilon_\nu \ll 10^{-6}$ is assume\textcolor{Navy}{d}, we do not worry about constraints from neutrinos. 
	}


	\section{Signal and Background from $J/\Psi \to \eta_c e^+ e^-$}
	\label{sec:spectrum}

{ In the SM, the decay $J/\Psi \to \eta_c \gamma^* \to \eta_c e^+ e^-$ is radiatively allowed~\cite{Gu:2019qwo}.
	 Its partial width is expressed with the form factor $f_{\rm VP}(0)$ for on-shell photon at the vanishing momentum transfer limit $q^2=0$~\cite{Fu:2011yy} 
	\begin{eqnarray}
	\Gamma(J/\Psi \to \eta_c \gamma) = \frac{1}{3}
	\frac{\alpha_{\rm EM} (m^2_{J/\Psi}-m^2_{\eta_c})^3}{8m^3_{J/\Psi}} |f_{\rm VP}(0)|^2,
	\end{eqnarray}
	from which the form factor $|f_{\rm VP}(0)|=0.68949~{\rm GeV^{-1}}$
	is determined with the fine structure constant $\alpha_{\rm EM}\simeq 1/128$, the masses of $J/\Psi$ and $\eta_c$, $m_{J/\Psi}=3.0969$ GeV, and $m_{\eta_c}=2.9839$ GeV, respectively, and the measured width $\Gamma(J/ \Psi \to \eta_c \gamma) = 1.5793  ~{\rm keV}$~\cite{Mitchell:2008aa, Anashin:2010nr, Becirevic:2012dc, Zyla:2020zbs}. The form factor $f_{\rm VP}(q^2)$ for general $q^2 \neq 0$ is obtained by $F_{\rm VP}(q^2)\equiv f_{\rm VP}(q^2)/f_{\rm VP}(0) 
	=1/(1-\frac{q^2}{\Lambda^2})$ from the pole approximation
	with pole mass $\Lambda=m_{\psi'}=3.686097$ GeV for $J/\Psi$.
}

{	The normalized differential widths for partial decay widths of $J/\Psi \to \eta_c \gamma^* \to \eta_c e^+ e^-$ and $J/\Psi \to \eta_c X^* \to \eta_c e^+ e^-$, respectively for off-shell photon and $X$ boson are obtained using a common factor $F_{\rm VP}(q^2)$
	~\cite{Gu:2019qwo}:
	\begin{align}
	\label{eq:dGam_dq2}
	\frac{d \Gamma_{\eta_c \gamma^*}}
	{dq^2 \Gamma_{J/\Psi \to \eta_c \gamma}}
	&=|F_{\rm VP}(q^2)|^2 \times F_{\rm QED}(q^2) \nonumber \\
	\frac{d \Gamma_{\eta_c X^*}}
	{dq^2 \Gamma_{J/\Psi \to \eta_c \gamma}}
	&=|F_{\rm VP}(q^2)|^2 \times F_X(q^2)\,,
	\end{align}
	where the kinematic window is given as $(2m_e)^2\leq q^2=m^2_{e^+e^-} \leq (m_{J/\Psi}-m_{\eta_c})^2$. 
Since the mass difference between $J/\Psi$ and $\eta_c$ is only 113 MeV,
it is hard to significantly boost $X$ boson in this channel and produce 
displaced-vertex signal in a collider detector.

	The precise expression for the factor $F_{\rm QED}$ is shown in Ref.~\cite{Gu:2019qwo} where the factor is found to include the amplitude square and phase space factor for off-shell photon. 
	Analogous expression for  $F_X$ is obtained:	
	\begin{eqnarray}
	F_X(q^2)& = & \frac{\alpha_{\rm EM} (\varepsilon_c \cdot \varepsilon_e)^2}
	{3\pi} 
	\left( \frac{q^2}{\left[(q^2-m^2_X)^2+m^2_X \Gamma^2_X \right]} \right) \nonumber \\
	&& \times
	\left(1-\frac{4m^2_e}{q^2} \right)^{1/2}
	\left(1+\frac{2m^2_e}{q^2} \right)
	\left[ \left(1+\frac{q^2}{m^2_{J/\Psi}-m^2_{\eta_c}} \right)^2
	- \frac{4m^2_{J/\Psi}q^2}{(m^2_{J/\Psi}-m^2_{\eta_c})^2} \right]^{3/2}
	\end{eqnarray}
	by replacing the couplings and propagator from $F_{\rm QED}$.
	}

	Assuming that $\varepsilon_\nu \ll 10^{-6}$ and $\varepsilon_e \simeq 10^{-3}$, and the quark channels are kinematically forbidden with $m_X \leq 2 m_\pi$,  the $X$ boson dominantly decay to electrons with the width
		\begin{eqnarray}
	\Gamma_{X \to e^+e^-} &=& \frac{\varepsilon^2_e \alpha_{\rm EM}m_X}{3}
	\left(1+\frac{2m^2_e}{m^2_X}\right) \sqrt{1-\frac{4m^2_e}{m^2_X}}\,,
	\end{eqnarray}
		which is narrow $\Gamma_X \ll m_X$.

	After performing the integration of $q^2$,
	we can obtain the partial decay width 
	$\Gamma_{\eta_c X^*}$.
	By inserting favoured coupling values 
	$\varepsilon_c= \varepsilon_u=3.7\times 10^{-3}$, $\varepsilon_e=10^{-3}$ 
	and fixing $m_X=17$ MeV for $^8{\rm Be}^*$ anomaly, 
	it gives $ {\Gamma_{\eta_c \gamma^*} 
	= 2.09\times 10^{-5}~{\rm keV}}$ and
	implies 
	$$
	{\rm Br}(J/\Psi \to \eta_c X^* \to \eta_c e^+ e^-) 
	= 1.64 \times 10^{-6}\, \left( \frac{\varepsilon_c}{10^{-2}} \right)^2\ ,
	$$ 
	which is about three orders of magnitude smaller than that of the $\eta_c \gamma^*$ background,
	${\rm Br}(J/\Psi \to \eta_c \gamma^* \to \eta_c e^+ e^-) 
	= 1.03\times 10^{-4}$.  
  \footnote{
  In the background estimation, the other background sources are not considered.
  For instance, $\gamma \to e^+ e^-$ conversion near the detector materials is contributed with $J/\Psi \to \eta_c \gamma$ decay as a background source.
  The process is suppressed by examining the vertex position of the resulting $e^+ e^-$. For the vertexing performance of Belle~II, the reader is directed to Ref.~\cite{Kou_2019}.
  }

	{The $e^+e^-$ invariant-mass-squared distributions for signal ($\eta_c X^*$) and background ($\eta_c \gamma^*$) are compared in  
	Fig.~\ref{fig:Jpsi}.  The different features are clearly displayed: the signal has a peak at $q^2 =m_X^2$ and the background is broadly distributed. Therefore our task now is to efficiently extract the signal near the peak and suppress the background.  
	}

	\begin{figure}[t!]
		\centering
		\includegraphics[height=3.4in,angle=270]{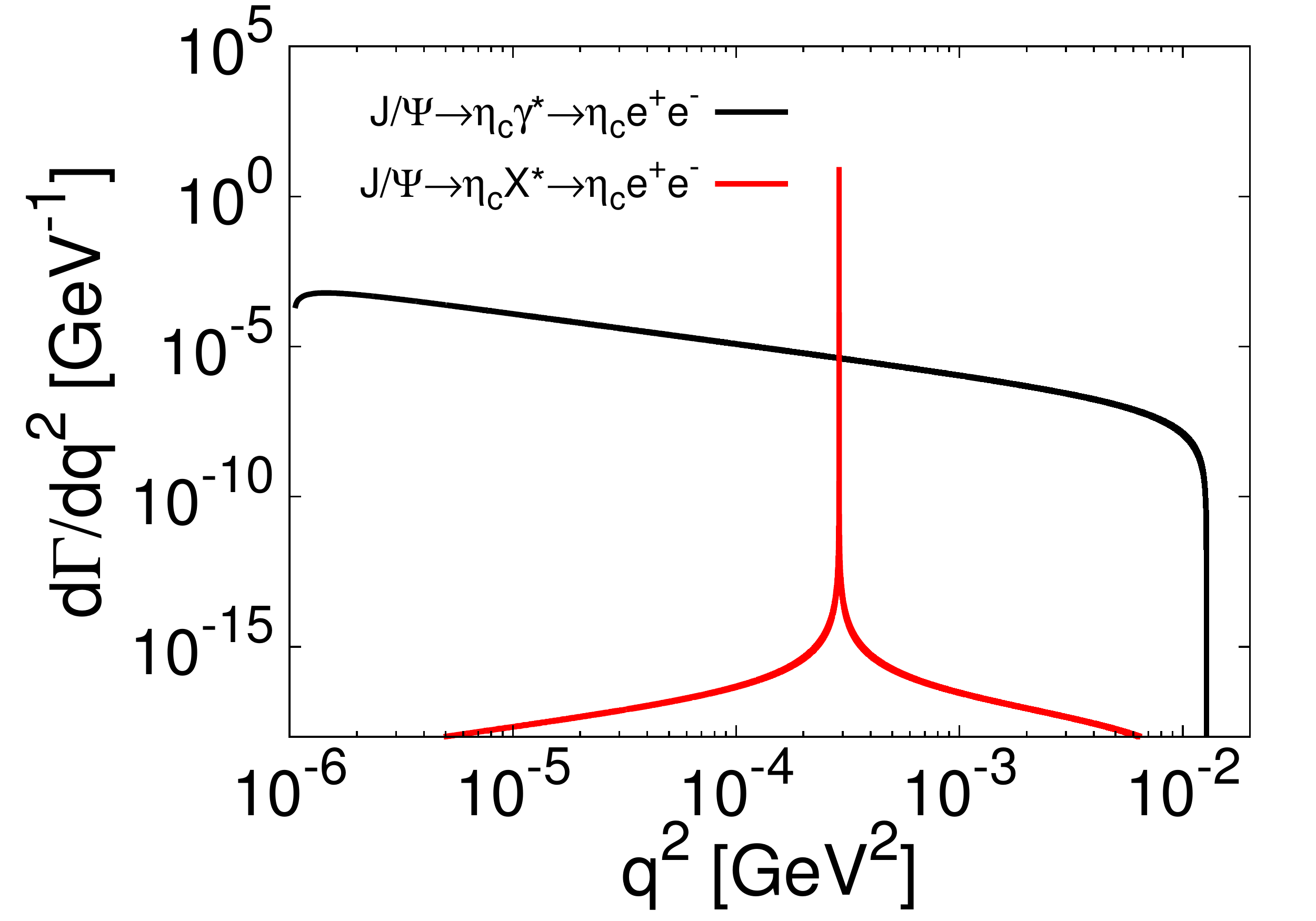}
		\caption{\small \label{fig:Jpsi}
			The $e^+e^-$ invariant mass distributions of 
			signal $J/\Psi \to \eta_c X^* \to \eta_c e^+ e^-$ (red)
			and background $J/\Psi \to \eta_c \gamma^* \to \eta_c e^+ e^-$ (black), where $q^2\equiv m^2_{e^+e^-}$.
			Input parameters are $m_X=17$ MeV, 
			$\varepsilon_c=3.7\times 10^{-3}$, 
			and $\varepsilon_e=10^{-3}$.
		}
	\end{figure}

  We first impose a kinematic condition for signal: 
  $M_{e^{+} e^{-}} \subset [(m_X-\sigma_m), (m_X+\sigma_m)]$, where $\sigma_m$ is the $e^+ e^-$ mass resolution which is roughly of the same order of magnitude as the energy resolution $\sigma_E$. 
	The signal yield $S$ and background yield $B$ are now obtained as 	
	\begin{eqnarray}
	S &=& N_{J/\Psi}\times 
	\frac{\int^{(m_X+\sigma_m)^2}_{(m_X-\sigma_m)^2} dq^2 
		\frac{d\Gamma_{\eta_c X^*}}{dq^2}}
	{\Gamma_{J/\Psi}} \,, \nonumber \\
	B &=& N_{J/\Psi}\times 
	\frac{\int^{(m_X+\sigma_m)^2}_{(m_X-\sigma_m)^2} dq^2 
		\frac{d\Gamma_{\eta_c \gamma^*}}{dq^2}}
	{\Gamma_{J/\Psi}} \,,
	\label{eq:SB}
	\end{eqnarray}
	where $N_{J/\Psi}$ is the total number of $J/\Psi$ produced in the collision, 
	and $\Gamma_{J/\Psi}=92.9$ keV is the total decay width of $J/\Psi$~\cite{Zyla:2020zbs}.

	The BESIII experiment, which has collected $10^{10}$ $J/\Psi$ events in the resonance process $e^+e^- \to J/\Psi$~\cite{Gu:2019qwo}, plans to increase the size of $J/\Psi$ sample to $10^{11}$ events in the near future. The energy resolution of BESIII is $\sigma_E/E\simeq 0.005$ for the final-state electron, which
	smears the invariant mass distribution of $e^+e^-$ by $\sigma_m \simeq 1$ MeV.
	
\begin{table}[t!]
	\caption{ \small \label{tab:eta_eff}
		The branching fractions of $\eta_c$ decay modes with corresponding efficiencies.
	}
	\begin{tabular}{cc}
		\hline
		Branching Ratio & Detection efficiency \\ \hline
		${\rm Br}(\eta_c \to K^+K^-\pi^0)=(1.15\pm 0.12)\%$ & 18.82\% \\ 
		${\rm Br}(\eta_c \to K^0_SK^\pm \pi^\mp)=(2.60\pm 0.21)\%$ & 21.22\% \\ 
		${\rm Br}(\eta_c \to 2(\pi^+\pi^-\pi^0))=(15.2\pm 1.8)\%$ & 3.07\% \\ \hline
	\end{tabular}
\end{table}	
  In order to exclusively reconstruct the $\JPsi \to \etacepem$ decays, we have to consider $\eta_c$ decay modes that can be fully reconstructed with reasonable background contamination. Table~\ref{tab:eta_eff} lists the branching fractions of a few such $\eta_c$ decay modes along with the corresponding efficiencies~\cite{Ablikim:2019ory}.

	The overall efficiency $\epsilon$ of the above three $\eta_c$ modes is obtained by adding the individual efficiencies weighted by their corresponding branching fractions: 
	$\epsilon = 1.23\%$.
  Given these, and taking 17 MeV $X$ boson for $^8{\rm Be}^*$ anomaly to be real, we list, in Table~\ref{tab:S_B}, the expected significances of $\JPsi \to \eta_c X \to \etacepem$  with $N_{\JPsi} = 10^{11}$ at BESIII  under the assumption of $\varepsilon_c=\varepsilon_u$, for various $\sigma_m$ values.
\begin{table}[h!]
	\caption{\small \label{tab:S_B}
		For $N_{J/\Psi}=10^{11}$ 
		and favoured parameters $\varepsilon_c= \varepsilon_u=3.7\times 10^{-3}$, $\varepsilon_e=10^{-3}$, $m_X=17$ MeV for $^8{\rm Be}^*$ anomaly, 
		the significances of signal to background from 
		$J/\Psi \to \eta_c e^+e^-$
		with various energy resolutions of detector 
		and $1.23\%$ $\eta_c$ reconstruction efficiency.
	}
	\begin{ruledtabular}
		\begin{tabular}{c|ccccc}
			& $\sigma_m$=1 MeV   & $\sigma_m$=2 MeV   & $\sigma_m$=5 MeV  & $\sigma_m$=10 MeV & $\sigma_m$=15 MeV  \\
			\hline
			$S$     
			& 188 & 263 & 277 & 277 & 277   \\
			$B$     
			& 3686 & 7399 & 18989 & 42436 & 87640    \\
			\hline
			$S/\sqrt{B}$    
			& 3.10 & 3.06 & 2.01 & 1.34 & 0.94 \\
		\end{tabular}
	\end{ruledtabular}
\end{table}

\begin{figure}[t!]
	\centering
	\includegraphics[height=1.8in,angle=0]{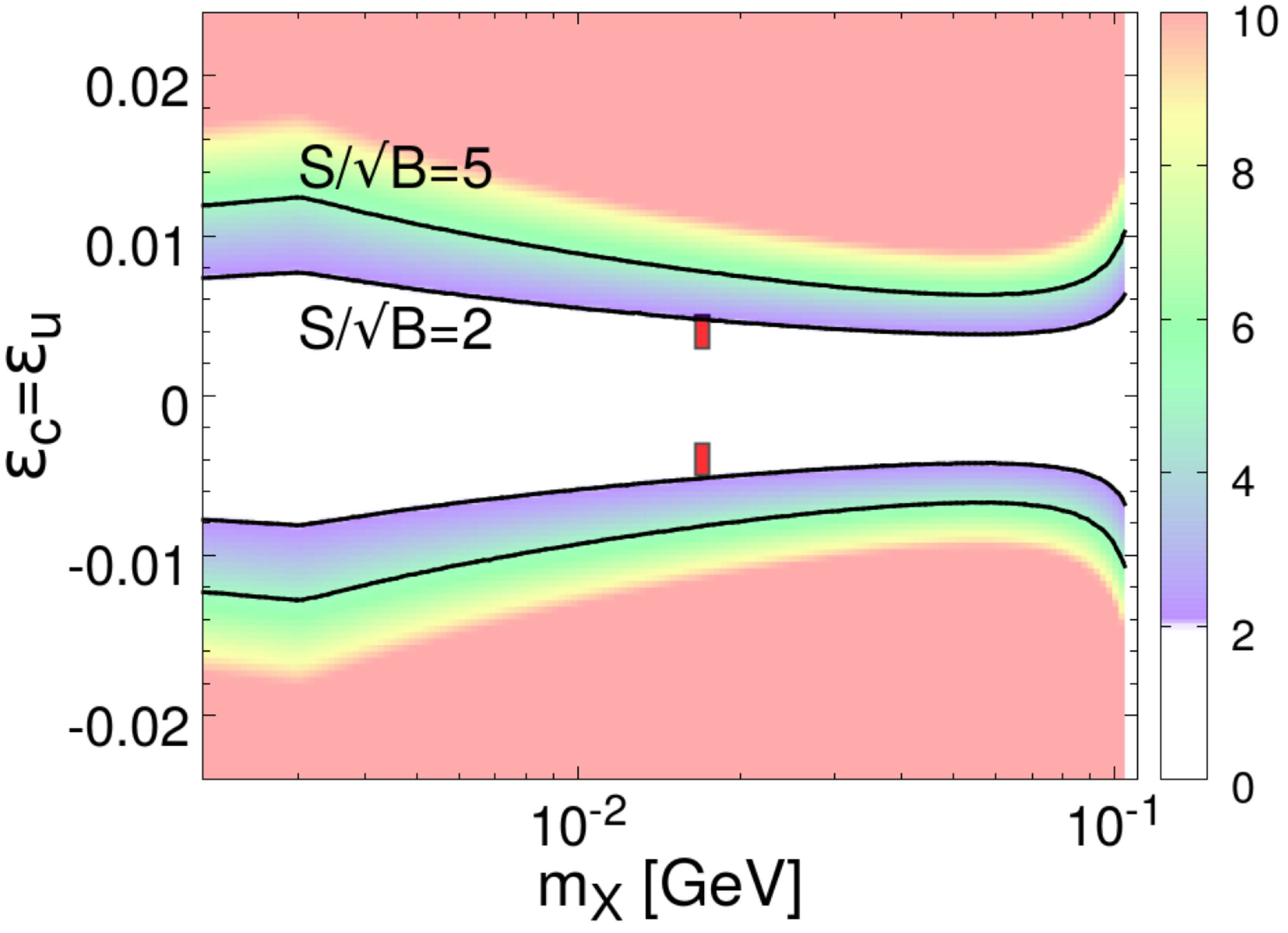}
	\includegraphics[height=1.8in,angle=0]{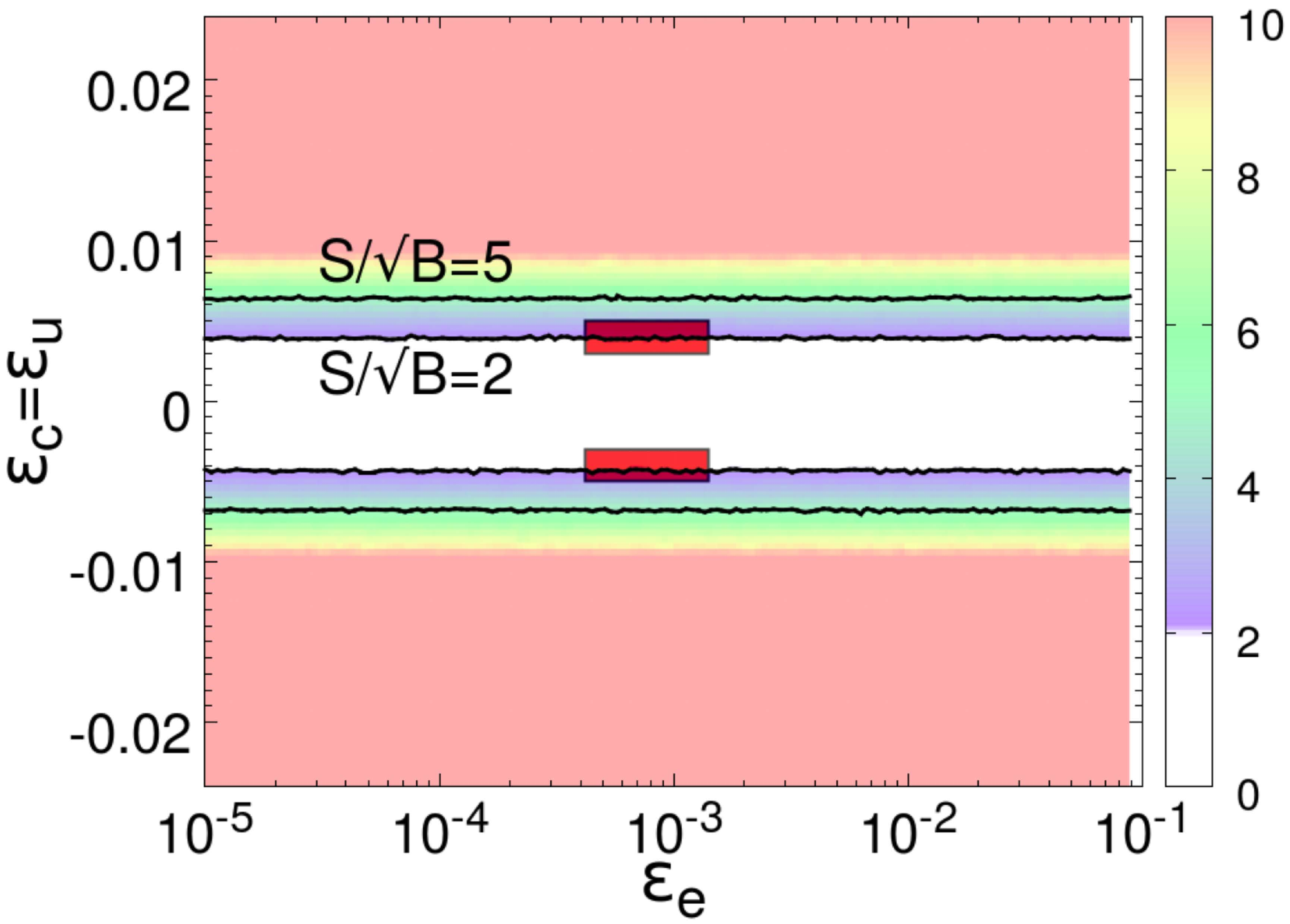}
	\includegraphics[height=1.8in,angle=0]{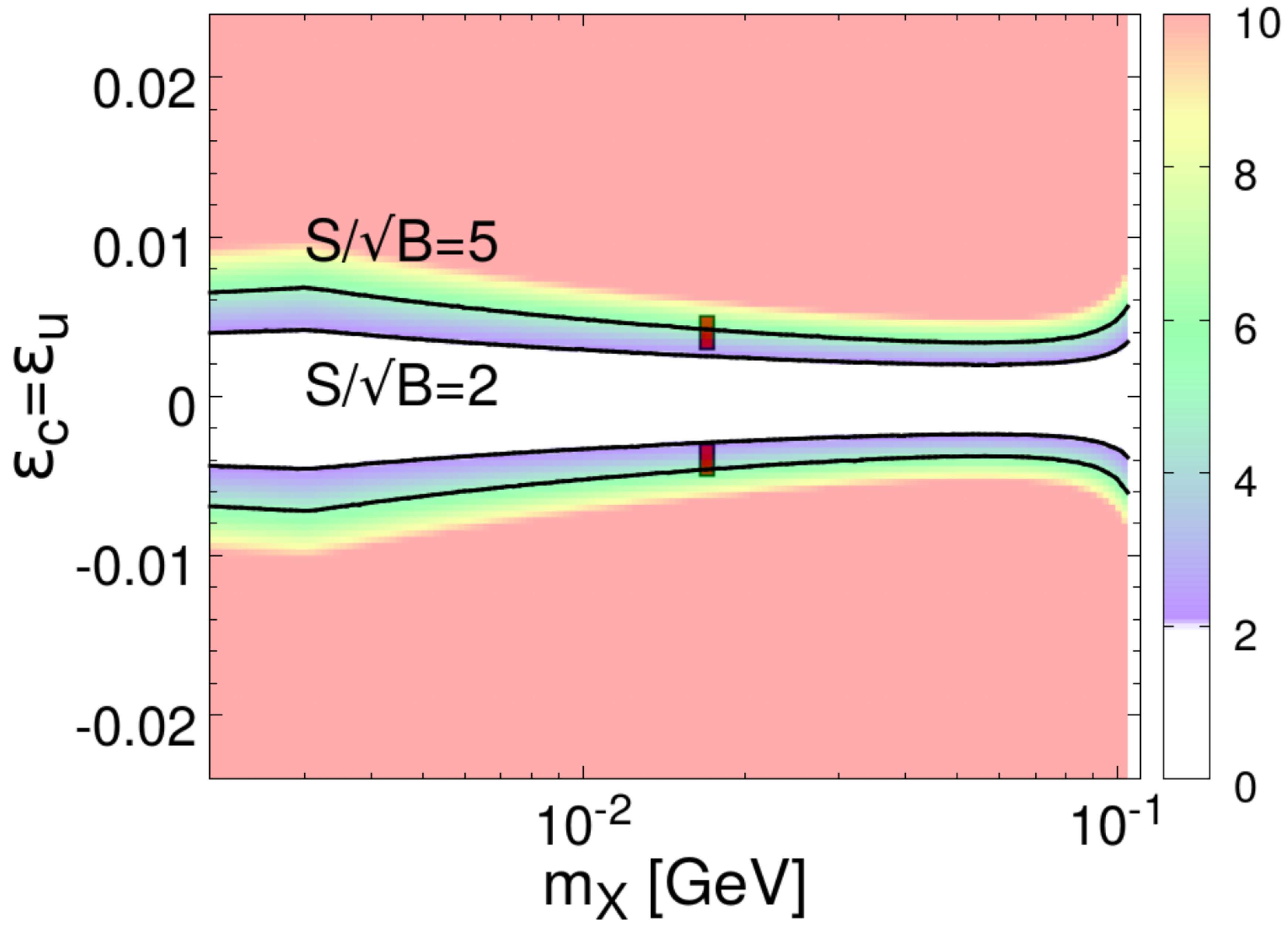}
	\includegraphics[height=1.8in,angle=0]{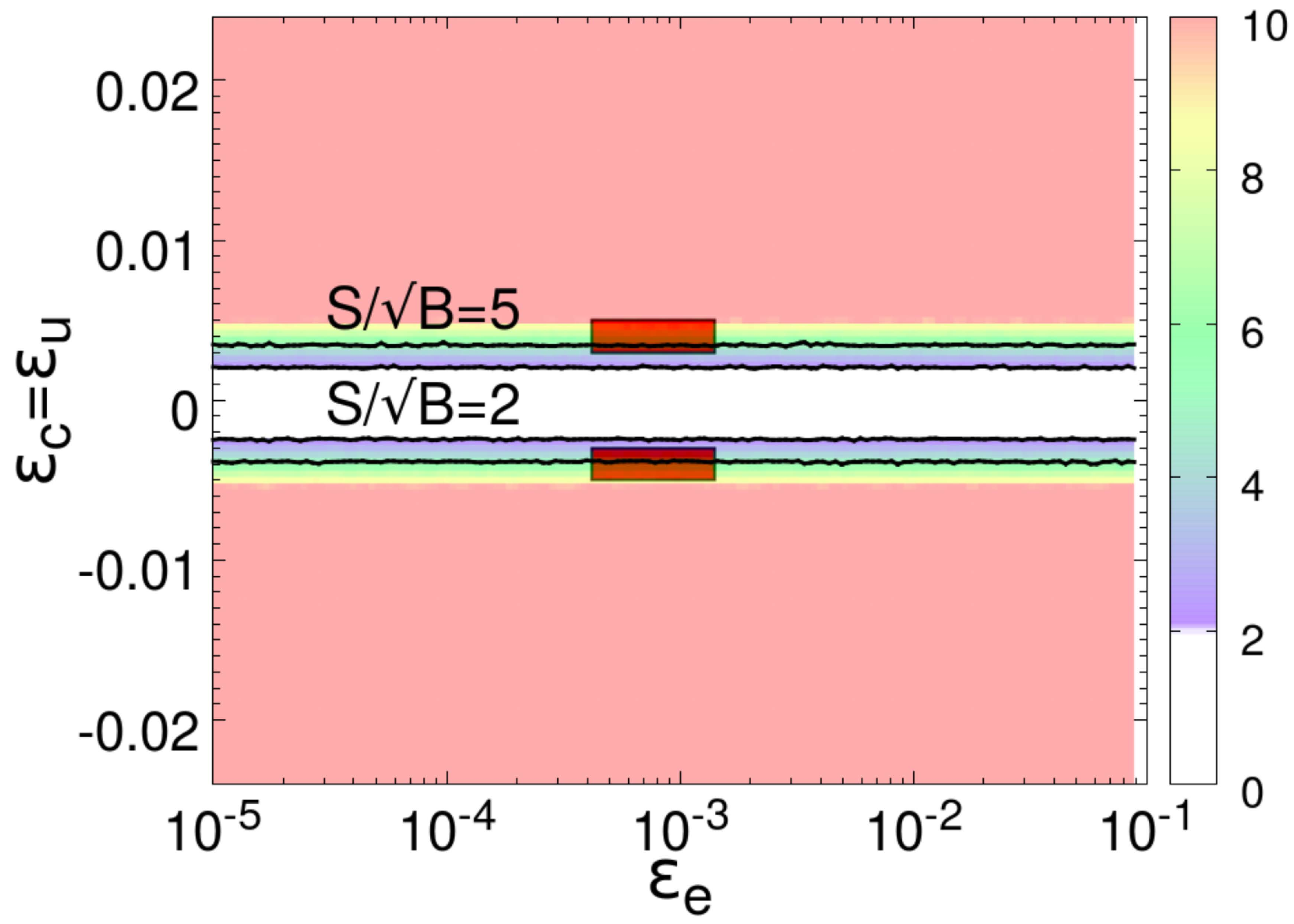}
	\caption{\small \label{fig:Jpsi_scan}
		The significance $S/\sqrt{B}$ on $(m_X,\varepsilon_c)$ and $(\varepsilon_c,\varepsilon_e)$ planes 
		from $J/\Psi \to \eta_c e^+e^-$ 
		light $X$ vector boson searches at BESIII.
		Adopt $N_{J/\Psi}=10^{10}$(upper panels) and $N_{J/\Psi}=10^{11}$(bottom panels), 
		reconstructed efficiency of $\eta_c$ from Table \ref{tab:eta_eff}, 
		and the invariant mass cut 
		$|M_{ee}-m_X|\leq \sigma_m=2\,{\rm MeV}$.
		The red boxes indicate the preferred regions for $^8{\rm Be}^*$ anomaly.
	}
\end{figure}

\begin{figure}[b!]
	\centering
	\includegraphics[height=1.8in,angle=0]{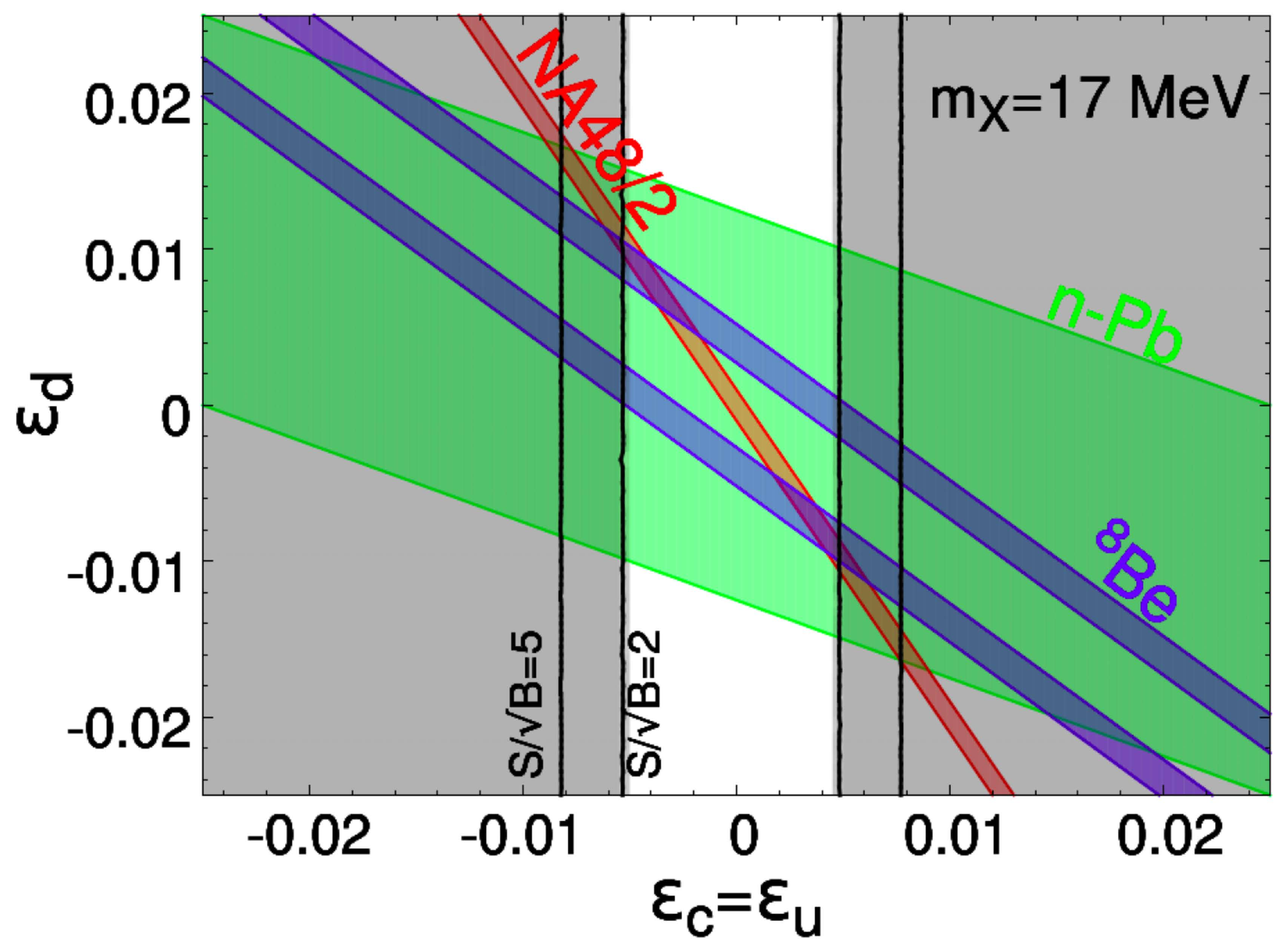}
	\includegraphics[height=1.8in,angle=0]{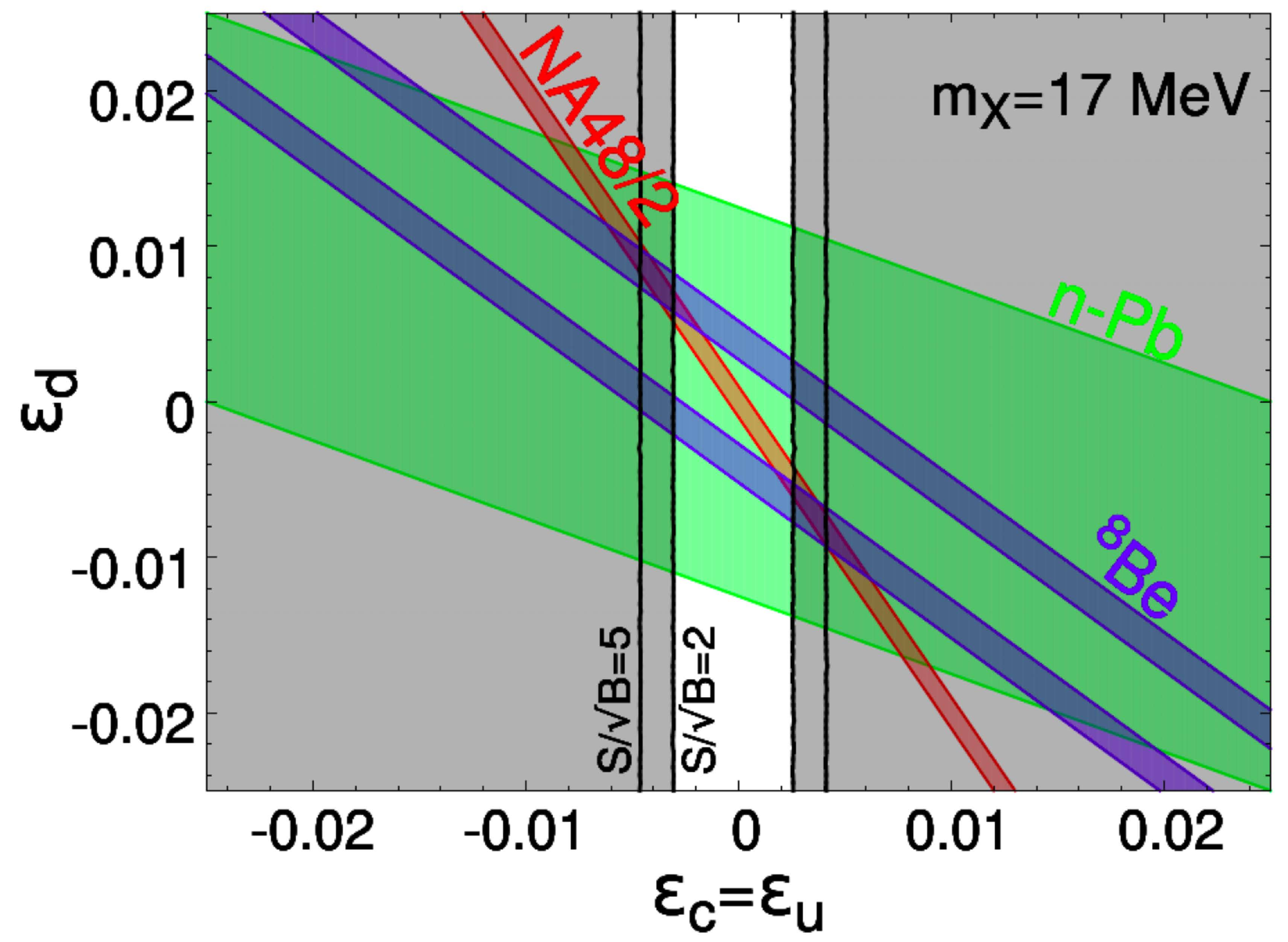}
	\caption{\small \label{fig:Jpsi_ideal}
		The current (left-panel: $N_{J/\Psi}=10^{10}$) and future (right-panel: $N_{J/\Psi}=10^{11}$) BESIII sensitivities
		by assuming the reconstruction efficiencies from 
		Table \ref{tab:eta_eff} and taking the invariant mass cut 
		$|M_{ee}-m_X|\leq \sigma_m=2\,{\rm MeV}$.
		Compare with the allowed regions for $^8{\rm Be}^*$ anomaly,
		NA48/2 for $\pi^0$ decay~\cite{Batley:2015lha} 
		and neutron-nucleus scattering~\cite{Barbieri:1975xy}.
	}
\end{figure}

	For general light vector boson searches through $J/\Psi \to \eta_c e^+e^-$,
	the variation of the expected significance over $(m_X,\varepsilon_c,\varepsilon_e)$ 
	are shown in Fig.~\ref{fig:Jpsi_scan}.
	With the present value of $N_{J/\Psi}=10^{10}$ at BESIII, the region of sensitivity
	is $|\varepsilon_c|\gtrsim 5 \times 10^{-3}$
	at $ m_X\simeq 17~{\rm MeV}$ as shown in
	the upper left panel of Fig.~\ref{fig:Jpsi_scan} and left panel of Fig.~\ref{fig:Jpsi_ideal}.
	The sensitivity slightly improves as $m_X$ increases,
	because of the reduction of background (see Fig.~\ref{fig:Jpsi}),
	and reaches the maximal sensitivity 
	$|\varepsilon_c|\ \gtrsim\ 3 \times 10^{-3}$
	at $ m_X\simeq 60~{\rm MeV} $.
	But as $m_X$ approaches $m_{J/\Psi}-m_{\eta_c}$, 
	the sensitivity becomes weaker due to 
	the phase space suppression.
	The two right panels of Fig.~\ref{fig:Jpsi_scan} show that the significance is independent of the $\varepsilon_e$ as we expect from the narrow width approximation.
	For $N_{J/\Psi}=10^{11}$ which is expected in the near future, 
  the projected sensitivity becomes $|\varepsilon_c|\ \gtrsim\ 3 \times 10^{-3}$
	at $ m_X\simeq 17~{\rm MeV}$ as shown in the bottom-left panel of 
	Fig.~\ref{fig:Jpsi_scan} and the right panel of Fig.~\ref{fig:Jpsi_ideal}, whereby the entire favored region of $^8{\rm Be}^*$ anomaly can be probed.

	An alternative way to explicitly reconstructing $\eta_c$ in $J/\Psi \to \eta_c e^+e^-$ at BESIII
	is to use the recoil of $e^+e^-$.
	As the $e^\pm$ carries low energy around 50 MeV,
	it gets difficult to distinguish $e^\pm$ tracks from $\pi^{\pm}$ background.
  With an improvement of low-energy electron identification in the future,
	the BESIII with $N_{J/\Psi}=10^{11}$ can
	reach the sensitivity of $|\varepsilon_c|\simeq 10^{-3}$.


	\section{The $e^+e^- \to \ell^+\ell^- + J/\Psi \to \ell^+ \ell^- e^+ e^- \eta_c$ at Belle~II}
	\label{sec:4lepton}
\begin{figure}[h!]
	\centering
	\includegraphics[height=1.8in]{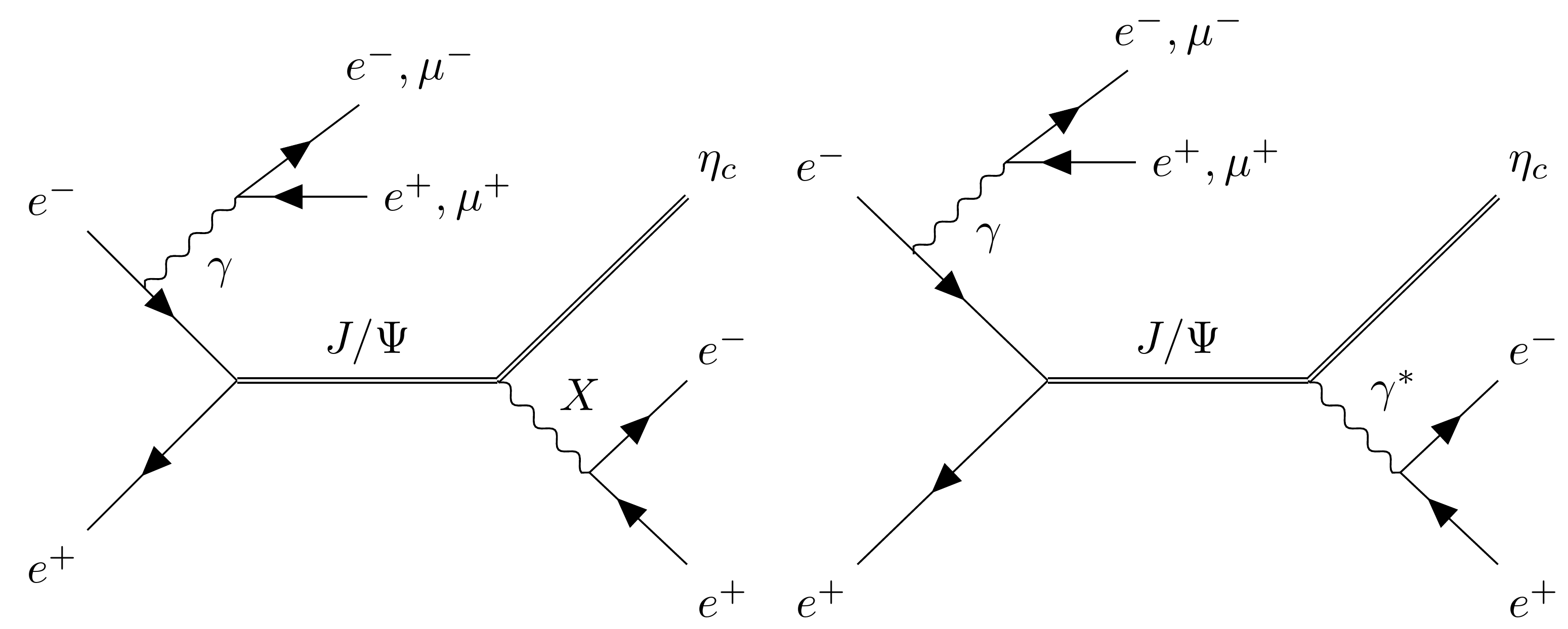}
	\caption{\small \label{fig:feyn}
		The feynman diagrams of the signal (left) and background (right)}
\end{figure}

	{
	For vector meson $J/\Psi$, the partial width to $e^+e^-$ is given by 
	\begin{eqnarray}
	\Gamma_{J/\Psi \to e^+ e^-}= \frac{g^2_{J/\Psi e e}}{12 \pi} m_{J/\Psi}
	\left( 1 + \frac{2m^2_e}{m^2_{J/\Psi}} \right)
	\sqrt{1-\frac{4 m^2_e}{m^2_{J/\Psi}}}\,,
	\label{eq:br}
	\end{eqnarray}
	where $g_{J/\Psi e e} =8.2048\times 10^{-3}$~\cite{Zyla:2020zbs} is the coupling strength
	in the effective interaction $g_{J/\Psi e e}\, [\bar{e}\gamma^\mu e] (J/\Psi)_\mu$ {that matches the measured value 
	$\Gamma_{J/\Psi \to e^+ e^-}=5.53~{\rm keV}$~\cite{Zyla:2020zbs}}. 
	
	Then the cross sections to $\ell^+ \ell^- J/\Psi$ where $\ell =e,$ or $\mu$ at Belle~II are obtained via $e^+e^-\to \gamma^* J/\Psi$: 
	\begin{align}
	& \sigma(e^+e^- \to \gamma^*+ J/\Psi \to e^+ e^- J/\Psi)=286~{\rm fb}, \nonumber \\
	& \sigma(e^+e^- \to \gamma^* + J/\Psi \to \mu^+ \mu^- J/\Psi)=58.4~{\rm fb}\,.
	\end{align}
	}

	With the design integrated luminosity ${\cal L}=50\,{\rm ab^{-1}}$, we estimate 
		$N_{\JPsi} = 1.75\times 10^7$ events for
	$e^+e^- \to \gamma^* + J/\Psi \to \ell^+ \ell^- J/\Psi$ at Belle~II. This $N_{\JPsi}$ is applied to Eq.~\ref{eq:SB}, along with Eq.~\ref{eq:br}, to give estimates of $S$ and $B$:
	\begin{eqnarray}
	\label{eq:s_b_event}
	&& S= {\cal L} \times 
	\sigma(e^+e^- \to  \ell^+ \ell^- J/\Psi)
	\times  {\rm Br}(J/\Psi \to \eta_c X^* \to \eta_c e^+e^-)
	\simeq 28.2 \left( \frac{\varepsilon_c}{10^{-2}} \right)^2,  \nonumber \\
	&& B= {\cal L} \times 
	\sigma(e^+e^- \to  \ell^+ \ell^- J/\Psi)
	\times  {\rm Br}(J/\Psi \to \eta_c \gamma^* \to \eta_c e^+e^-)
	\simeq 1772.
	\end{eqnarray}
	{Therefore, the estimated $S/\sqrt{B}$ is too small at this level so that we will improve the analysis by
  a more realistic MC study below. }


	\label{event_selectoion}

	\begin{figure}[t!]
		\centering
		\includegraphics[height=2.0in]{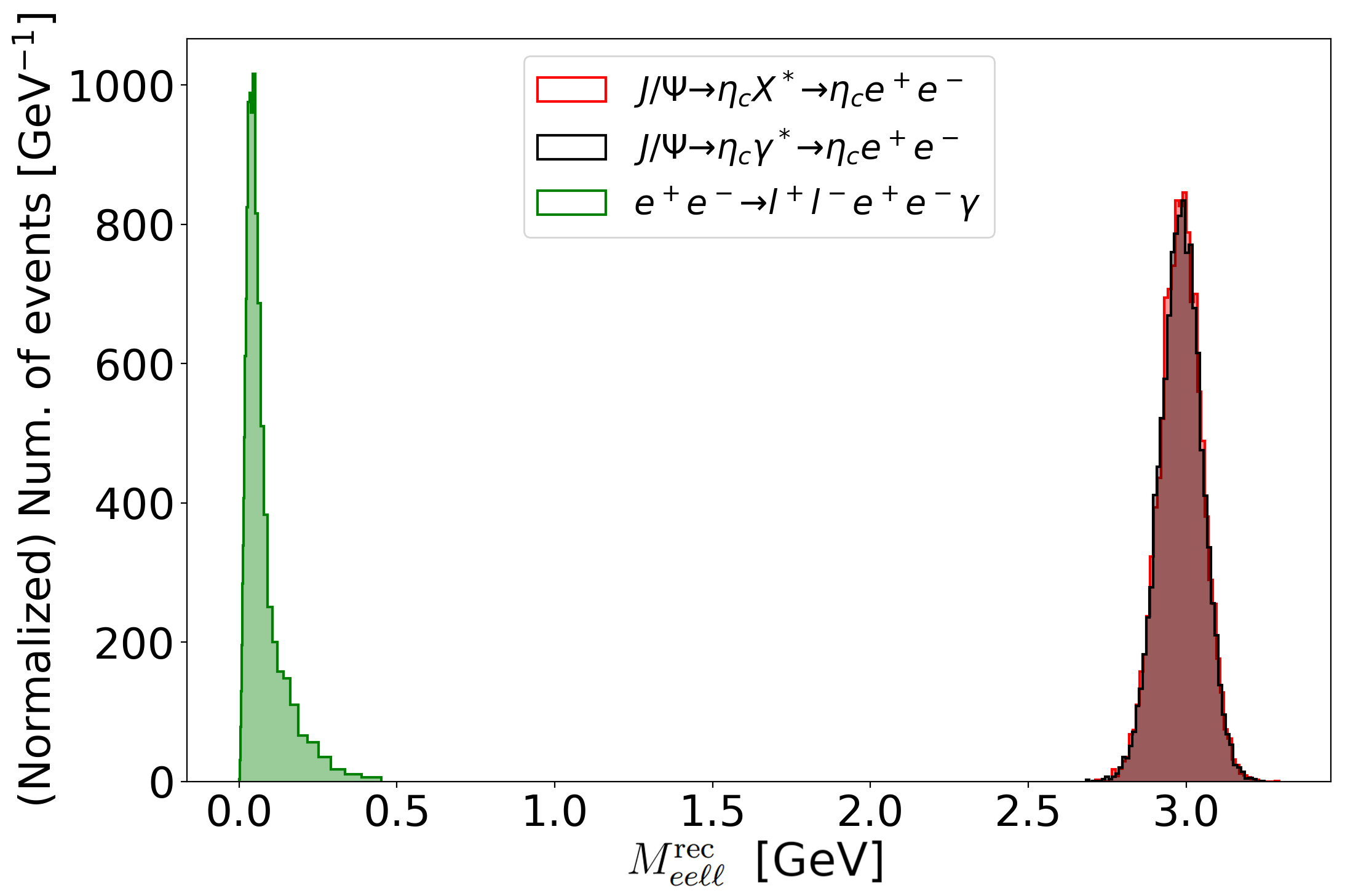}
		\includegraphics[height=2.0in]{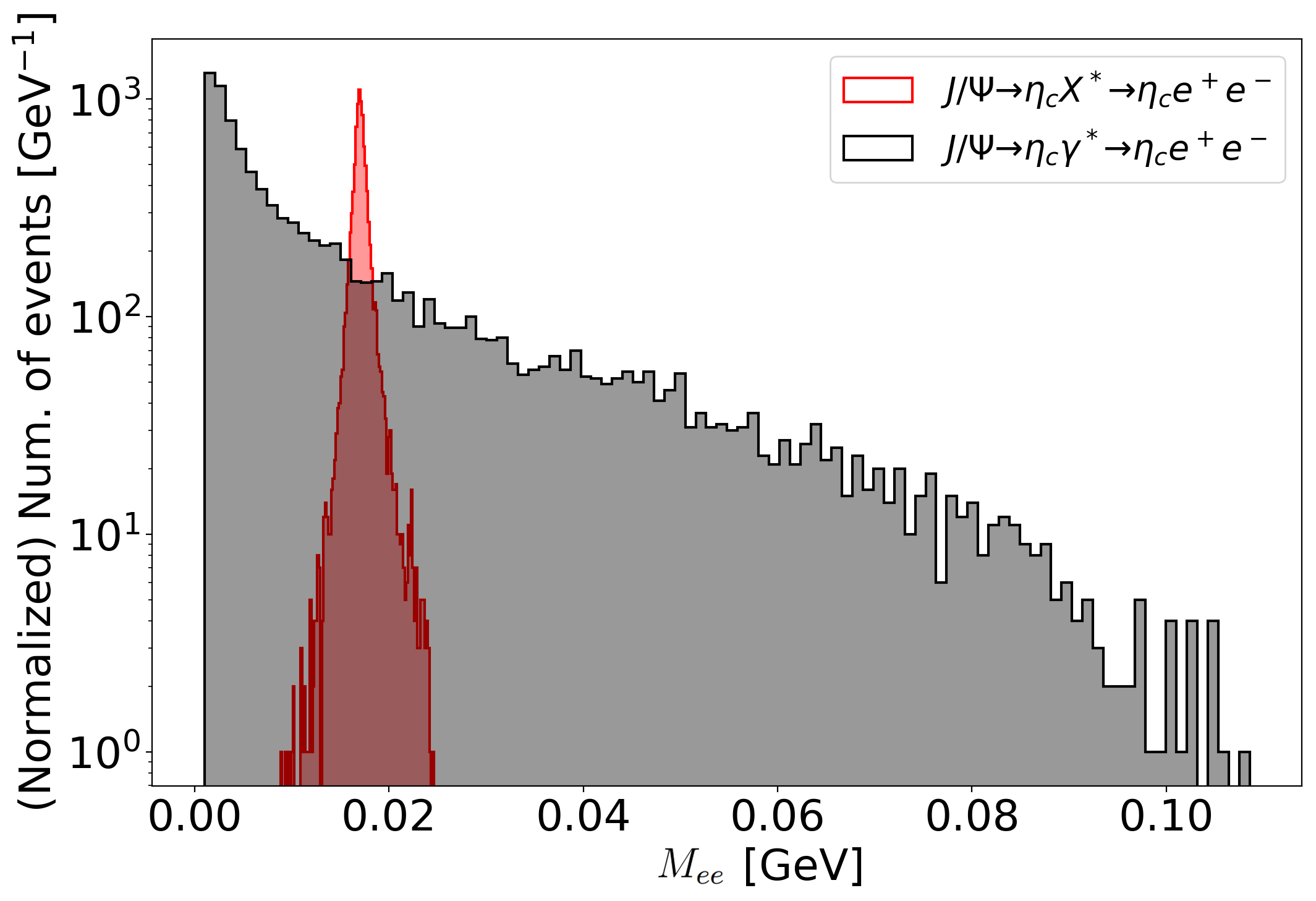}
		\caption{\small \label{fig:Jpsi_mc}
			The $e^+e^-\ell^+\ell^-$ recoil mass (left) and $e^+e^-$ invariant mass (right)  distributions
			for the parton level Monte-Carlo simulation data with the smearing effect.
			Input parameter is $m_X=17$ MeV.
			Here, we normalized to $10^5$ events for each channel.
		}
	\end{figure}

	For event generation, we use \textsc{MG5\_aMC@NLO} \cite{Alwall:2014hca} for both signal and background with \textsc{FeynRules v2.0} \cite{Alloul:2013bka} model for $J/\Psi, \eta_c$ mesons and $X$ boson couplings and $X$ couples to the leptons.  We generate with the $E_{\mathrm{beam1,2}}=5.2941$ GeV in the CM frame, which is boosted by $\beta=0.2732$ with respect to the lab frame. 
	The amplitude of the electromagnetic Dalitz decay, $\VtoPepem$ can be written in a Lorentz-invariant form \cite{Gu:2019qwo},
	\begin{equation}
	T(\VtoPepem)=4\pi \alpha_{\rm EM} f_{\rm VP}\epsilon^{\mu\nu\rho\sigma}p_\mu q_\nu \epsilon_\rho \frac{1}{q^2}\bar{u}_1 \gamma_\sigma \nu_2
	\end{equation}
	and we can obtain the interaction Lagrangian as,
	\begin{equation}
	\label{eq:Lag_VPee}
	\mathcal{L} \supset f_{\rm VP} ( -2\sqrt{\pi \alpha_{\rm EM}}\partial_\mu P \partial_\nu V_\rho \epsilon^{\mu\nu\rho\sigma}A_\sigma- g_{Xc}  \partial_\mu P \partial_\nu V_\rho \epsilon^{\mu\nu\rho\sigma} X_\sigma ) - g_{eV} \bar{e} \gamma^\mu e V_\mu - g_{Xe}\bar{e} \gamma^\mu e X_\mu ~, 
	\end{equation}
	{where $g_{Xc}$, $g_{Xe}$, and $g_{eV}$ are the effective coupling constants, whose numerical values are to be obtained by experiments.}
	
	{
		Similar to the BESIII, photon conversion { process where the photon from $J/\Psi \to \eta_c \gamma$ hits the beam pipe or vertex detector layers and converts to $e^+ e^- $ can be a background source, but they are controlled by examining the $e^+ e^-$ vertex position.} Another possible background is  $e^+ e^- \to \ell^+ \ell^- + (\mathrm{anything})$ where (anything) may contain a number of low-momentum charged particles that are misidentified as electrons.  However, this process { does not show peak in the $M_{\ell^+ \ell^-}^{\mathrm{recoil}} \simeq M_{J/\Psi}$ variable, and can be suppressed by inspecting the $M_{\ell^+ \ell^-}^{\mathrm{recoil}}$ distribution.} In the Fig.~\ref{fig:Jpsi_mc}, {the recoil mass of $\epem\ell^+\ell^-$ and invariant mass distributions of $\epem$} are plotted for the signal (shown in red), and {the $\ell^+\ell^-e^+e^- (\gamma)$ (green) and $J/\Psi \to \eta_c \gamma^*$ (gray) backgrounds}, { where we produce equal number of events ($=10{^5}$) for each sample at the parton level before applying any selection cuts.} 
We give a Gaussian smearing effect with the momentum resolution $\sigma_{p_{\ell^\pm}}/p_{\ell^{\pm}}=0.005$ on the parton level data for our analysis.	
	To simulate the effects of the Belle~II detector, we apply the following baseline cuts: $|\eta^*_{\ell^\pm}| \leq 1.60$ in the CM frame~\cite{Jho:2019cxq, Adachi:2019otg}, $|E_{\mu^\pm}|\geq0.6$ GeV, and $|E_{e^\pm}|\geq0.06$ GeV in the lab frame \cite{Kou:2018nap}. 
{ Note that we require very low energy threshold for electrons so as to keep most of the signals, because the $e^+e^-$ from $J/\Psi \to \eta_c e^+e^-$ are very soft. This inevitably would cause large background from pion tracks being misidentified as electrons.\footnote{ Currently, the most powerful observable for electron identification ($e$-ID) at Belle~II is the $E/p$ where $E$ is the energy measured in the electromagnetic calorimeter and $p$ is the magnitude of 3-momentum measured in the drift chamber.~\cite{HANAGAKI2002490} With low-momentum tracks that cannot reach the calorimeter~\cite{Bertacchi_2021}, the performance of $e$-ID will thus degrade. For the actual data analysis, we encourage the Belle~II collaboration to improve $e$-ID for low-momentum tracks along with systematic uncertainties.}}
 
	{And we give two kinematic requirements:} For the energy-momentum 4-vector of $\eta_c$, we use the energy and momentum recoiling against $\epem \lplm$. 
	The signal and background distributions of the recoil mass $\Mreceell$, smeared by the charged-track momentum resolution, are displayed in the {left} panel of Fig.~\ref{fig:Jpsi_mc}.
	{The $J/\Psi \rightarrow \eta_c \epem$ events clearly show a peak at $\Mreceell \simeq m_{\eta_c}$, while the $\ell^+\ell^-e^+e^- (\gamma)$ background is mostly populated at $\Mreceell \simeq 0$,
	therefore} we require $|\Mreceell-m_{\eta_c}| \leq 200~{\rm MeV}$ to eliminate the $\ell^+\ell^-e^+e^- (\gamma)$ background. 
	In addition, we apply $|M_{ee}-m_X|\leq 2~{\rm MeV}$, whereby the $\eta_c \gamma^*$ background from the process shown in the right panel of Fig.~\ref{fig:feyn} are suppressed.
	In Table~\ref{tab:llee_S_B}, we summarize the cumulative effects of the baseline cuts and the kinematic requirements.}
	
\begin{table}[t!]
	\caption{\small \label{tab:llee_S_B}
		Signal and background events of $e^+e^- \to \ell^+\ell^- e^+e^- \eta_c$ after cuts at Belle~II.
	}
	\begin{adjustbox}{width=10cm}
		\begin{tabular}{c|c|c}
			\hline
			Cuts & \multicolumn{1}{c|}{B} & S \\
			\hline
			Processes    &~~ $\eta_c \gamma^* \to \eta_c ee$ ~~ 
			&~~ $\eta_c X \to \eta_c ee$ ~~ \\
			\hline
			& 100000 & 100000 \\
			Baseline Cuts    & 7170  & 6290 \\			
			$|\Mreceell-m_{\eta_c}| \leq 200~{\rm MeV}$ &7071 & 6219\\
			$|M_{ee}-m_X|\leq 2~{\rm MeV}$ & \textbf{377} &\textbf{5880}\\
			\hline
			\hline
		\end{tabular}
	\end{adjustbox}
\end{table}

	The sensitivities of $\ell^+\ell^-e^+e^-$ search at Belle~II
	are derived from the requirement of $S/\sqrt{B}=2$.
	Combining  Eq.~(\ref{eq:s_b_event}) and Table~\ref{tab:llee_S_B},
	we obtain the corresponding values of $\varepsilon_c$ with respect to luminosities of
	$50,100$, and $200~{\rm ab^{-1}}$ in Table~\ref{tab:4l_belleII}.
		They are about 5 times larger than the estimates from current BESIII sensitivity
		$|\varepsilon_c| \gtrsim 5 \times 10^{-3}$ in Section~\ref{sec:spectrum}.

	\begin{table}[h!]
		\caption{\small \label{tab:4l_belleII}
			Sensitivities on $\varepsilon_c$ of 17 MeV $X$ boson 
			from $\ell^+\ell^-  J/\Psi \to \ell^+ \ell^- e^+ e^- \eta_c$ search at Belle~II with luminosities 50, 100, and 200 ${\rm ab^{-1}}$.
			Here we require $S/\sqrt{B}=2$.
		}
		\begin{ruledtabular}
			\begin{tabular}{c|ccc}
				Luminosity & 50 ${\rm ab^{-1}}$   & 100 ${\rm ab^{-1}}$   & 200 ${\rm ab^{-1}}$  \\
				\hline
				$|\varepsilon_c|$  & $\gtrsim 1.76\times 10^{-2}$ & $\gtrsim 1.48\times 10^{-2}$ & $\gtrsim 1.24\times 10^{-2}$   \\
			\end{tabular}
		\end{ruledtabular}
	\end{table}

	\section{The $e^+e^- \to X + J/\Psi \to e^+e^- + J/\Psi $ Displaced Vertex at Belle~II}
	\label{sec:displaced}

	{The $X$ boson produced in the process  $e^+e^- \to X + J/\Psi$ 
	 travels several millimeters before decaying into $e^+e^-$ in the Belle~II detector and leaves displaced vertex.
	In particular, when the distance of the flight is between 2~mm and 8~mm, which is inside the beam pipe, 
	and outside the interaction region, Belle~II has excellent power to reconstruct the displaced vertex 
  and makes the signal almost free from the SM background~\cite{Duerr:2019dmv}.
	Therefore, we propose to use the clean displaced $e^+e^-$ 
	vertex from the $X$ boson decay (along with prompt $\ell^+\ell^-$ from $J/\Psi$).} 
	
\begin{figure}[t!]
	\centering
	\includegraphics[height=2.1in]{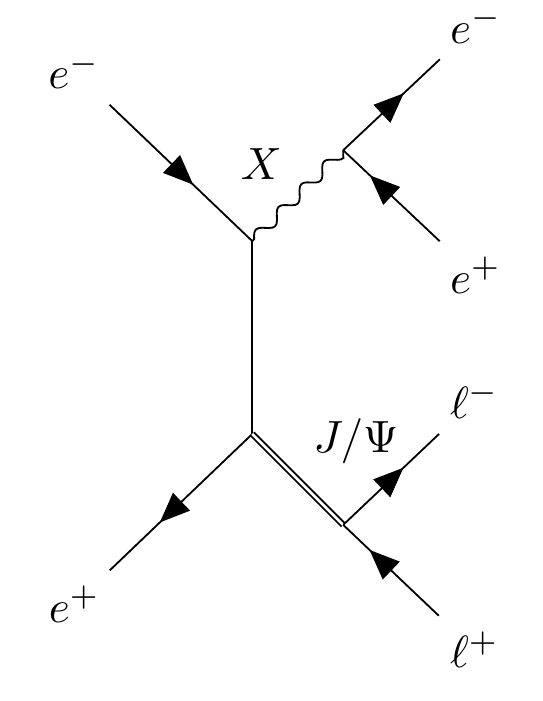}
	\hspace{2cm}
	\includegraphics[height=2.1in]{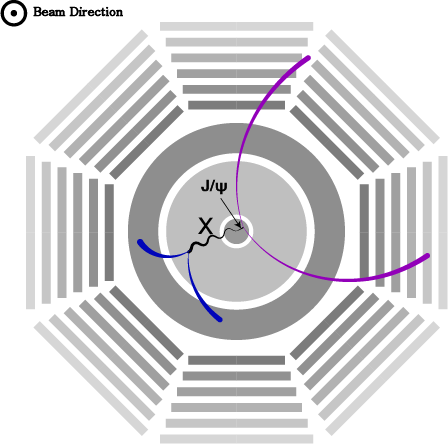}
	\caption{\small \label{fig:feyn_XJpsi}
		{\bf (Left Panel)} The Feynman diagram of the 
		$e^+e^- \to X+ J/\Psi \to e^+e^- \ell^+\ell^-$ signal
		{\bf (Right Panel)} The schematic picture of the decay process 
		of the long-lived $X$ boson in the Belle~II detector.}
\end{figure}

{	The leading-order Feynman diagram of the relevant process is shown in the left panel of Fig.~\ref{fig:feyn_XJpsi}. A typical event with displaced vertex at Belle~II detector with the $\ell^+\ell^-$ from $J/\Psi$ decay is schematically shown in the right panel of Fig.~\ref{fig:feyn_XJpsi}.
	}

	{We note that compared with other lighter vector mesons,
	the heavier mass of $J/\Psi$ helps to induce larger scattering angle
	in such a way that more events from $X \to e^+e^-$ will satisfy the cut $|\eta^*_{\ell^\pm}|\leq 1.60$.  } 
	Furthermore, the electron and positron from $X \to e^+e^-$ 
	carry energy above GeV, which make them easier to be distinguished 
	from charged-pion backgrounds.

	The {other advantage of this channel is that the signal strength only depends on the $\varepsilon_e$ coupling since only $X$-$e^+$-$e^-$ vertices are involved at tree level.} 
	For $0.3\times 10^{-3}\leq \varepsilon_e \leq 0.8\times 10^{-3}$, 
	it yields a few mm transverse flight distance $d_{xy}$ which is defined as the distance between the beam axis and the $X$ decay vertex.
	The left-panel of Fig.~\ref{fig:eps_dxy} shows the distribution of $d_{xy}$
	corresponding to several values of $\varepsilon_e$. 
	
	\begin{figure}[h!]
	\centering
	\includegraphics[height=1.8in]{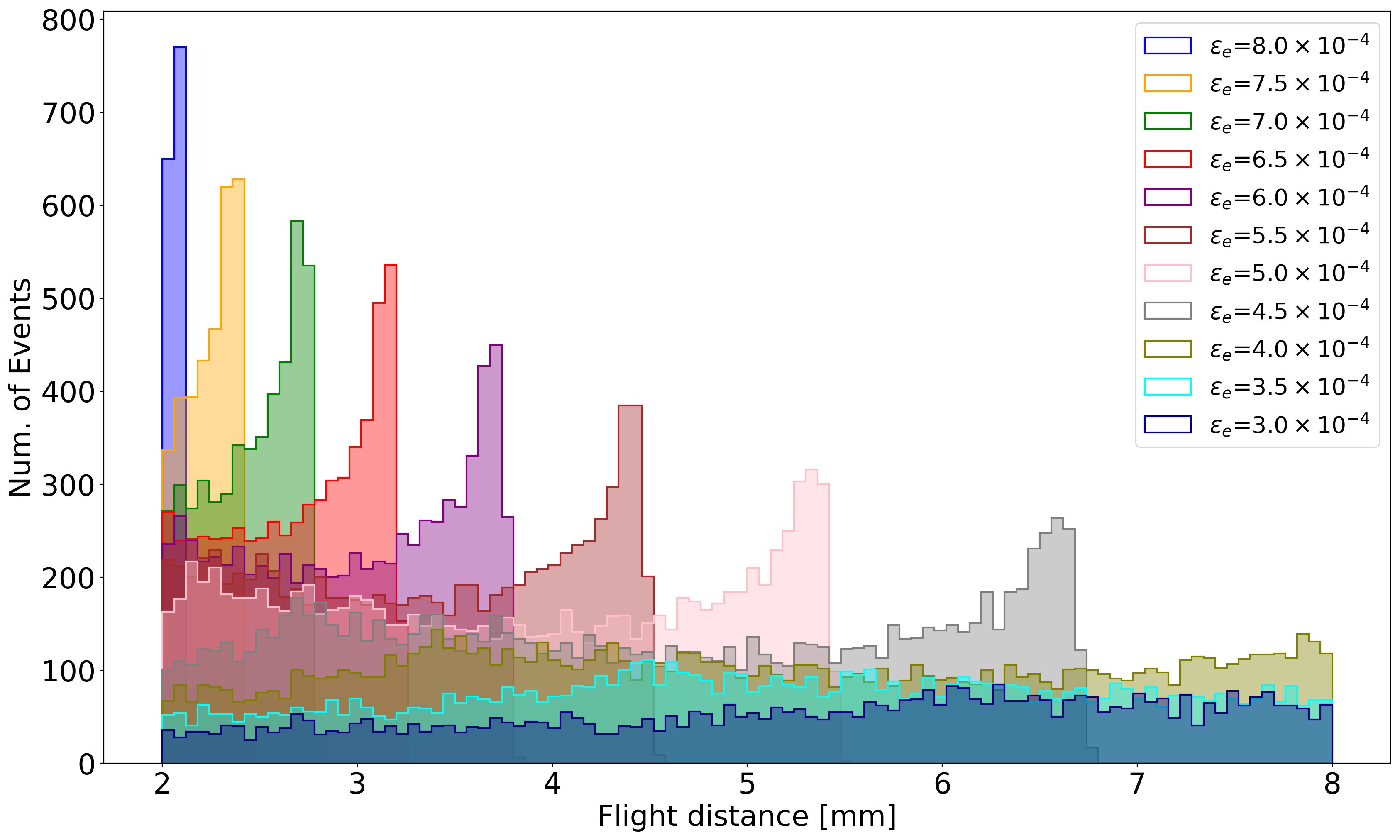}
	\includegraphics[height=1.8in]{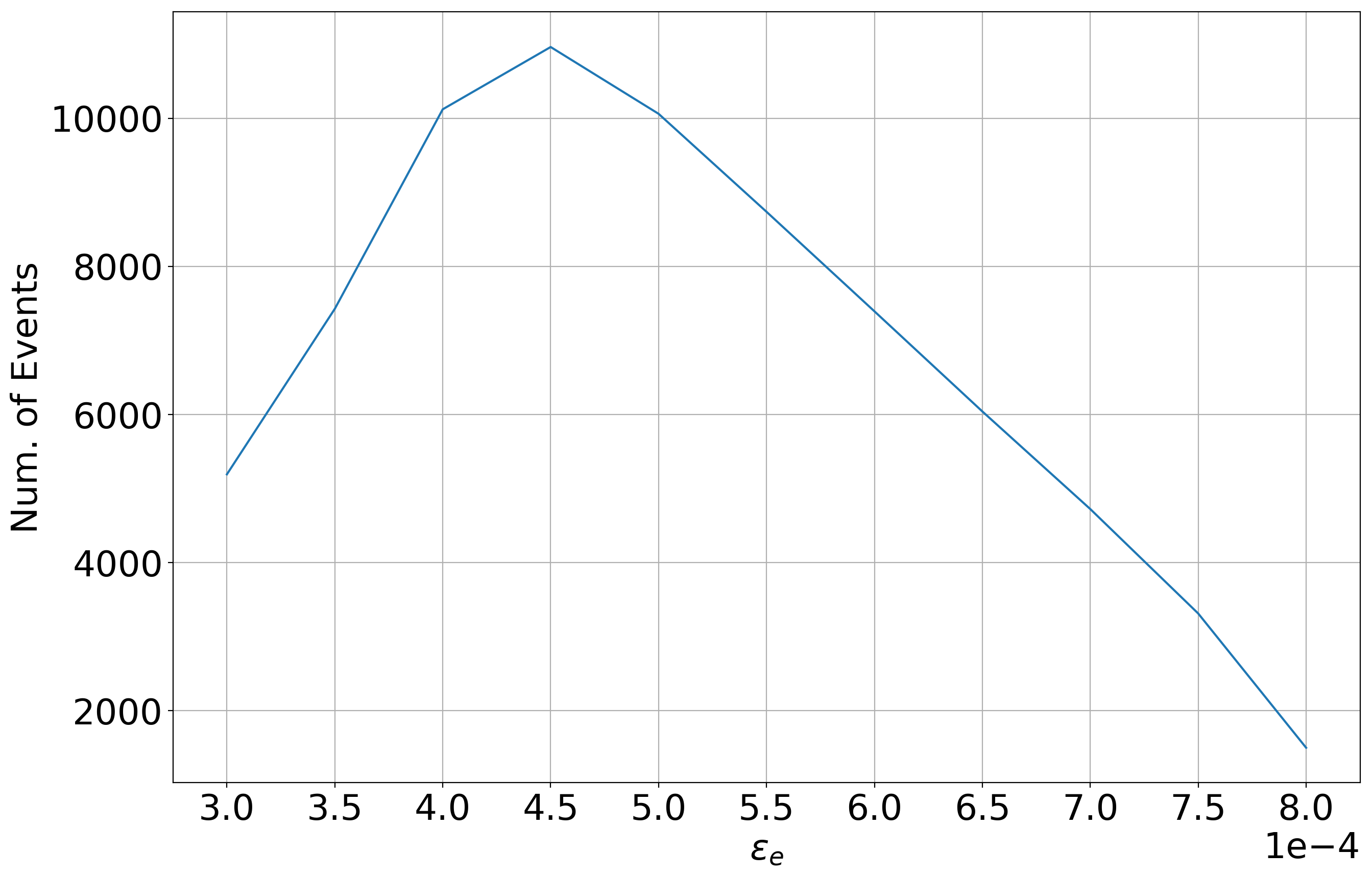}
	
	\caption{\small \label{fig:eps_dxy}
		(Left panel) The transverse flight distance $d_{xy}$ of 17 MeV $X$ boson for $\varepsilon_e$ in 0.5$\times 10^{-4}$ steps from 3$\times 10^{-4}$ to 8 $\times 10^{-4}$; 
		(Right panel) The number of events that pass the baseline cuts and satisfy 
		$2~{\rm mm}\leq d_{xy} \leq 8~{\rm mm}$ from the sample of $10^5$ generated signal events.}
	\end{figure}

\begin{table}[t!]
	\caption{\small \label{table:eps_dxy} {(From top rows to bottom)
		The fraction of events (in \%) that survive the baseline cuts mentioned in Section.\ref{event_selectoion}, and the flight distance $2~{\rm mm}\leq d_{xy} \leq 8~{\rm mm}$ cuts, for several values of $\varepsilon_e$ of 17 MeV $X$ boson;
		the number of signal events $N_S$ in the
		$e^+e^- \ell^+\ell^-$ channel at Belle~II with 
		$50~{\rm ab^{-1}}$ luminosity;
		and the expected significances assuming 1 event, and 0.1 event of the SM background in the analysis channel after all cuts.}}
	\begin{tabular}{|c||C{1.5cm}|C{1.5cm}|C{1.5cm}|C{1.5cm}|C{1.5cm}|C{1.5cm}|C{1.5cm}|}
		\hline\hline
		$\varepsilon_e/10^{-4}$ & $8.0$ & $7.0$ & $6.0$ & $5.0$ & \textbf{$4.5$} & $4.0$ & $3.0$ \\ \hline
		Baseline Cuts(\%) & 13.8 & 13.8 & 13.8 & 13.8 & 13.8 & 13.8 & 13.8   \\ \hline
		2mm $< d_{xy} <$ 8mm (\%) & 1.5 & 4.7 & 7.4 & 10.1 & 11.0 & 10.1 & 5.2   \\ \hline
		$N_S$  & 1.60  & 3.85 & 4.42 & 4.18 & 3.69 & 2.69 & 0.78   \\ \hline
		 Significance ($B=0.1$)  & $2.4\sigma$ & $4.6\sigma$ & $5.0\sigma$ & $4.8\sigma$ & $4.5\sigma$ & $3.6\sigma$ & $1.5\sigma$   \\ \hline
		 Significance ($B=1$)  & $1.6\sigma$  & $2.9\sigma$ & $3.2\sigma$ & $3.1\sigma$ &  $2.8\sigma$ & $2.3\sigma$  & $1.2\sigma$   \\ \hline\hline
	\end{tabular}
\end{table}
\begin{table}[t!]
	\caption{\small \label{table:eps_dxy_1}
		The same as Table~\ref{table:eps_dxy}, 
		but using the $e^+e^-$ channel}
	\begin{tabular}{|c||C{1.5cm}|C{1.5cm}|C{1.5cm}|C{1.5cm}|C{1.5cm}|C{1.5cm}|C{1.8cm}|}
		\hline\hline
		$\varepsilon_e/10^{-4}$ & $8.0$ & $7.0$  & $5.0$ & $4.0$ & $3.0$ & $2.0$ & $1.0$ \\ \hline
		Baseline Cuts(\%) & 17.6 & 17.6  & 17.6  & 17.6 & 17.6 & 17.6 & 17.6   \\ \hline
		2mm $< d_{xy} <$ 8mm(\%) & 1.6 & 5.3  & 12.3  & 12.9 & 7.4 & 2.3 & 0.5   \\ \hline
		$N_S$  & 14.6  & 35.7  & 42.7  & 28.7 &  9.23 & 1.28 &  0.07  \\ \hline
		{Significance ($B=0.1$)}  & \multicolumn{5}{c|}{$>5\sigma$} & $2.2\sigma$  & $0.4\sigma$   \\ \hline
		{ Significance ($B=1$)}  & \multicolumn{5}{c|}{$>5\sigma$} & $1.6\sigma$ & $0.9\sigma$   \\ \hline\hline
	\end{tabular}
\end{table}	

	With the baseline cuts and $2~{\rm mm}\leq d_{xy} \leq 8~{\rm mm}$, 
	we estimate the signal sensitivity by considering two cases: (i) explicitly reconstructing $J/\Psi \to \ell^+ \ell^-$ (`$e^+e^- \ell^+\ell^-$ channel'), and (ii) using the recoil mass of $X \to e^+ e^-$ to infer $J/\Psi$ (`$e^+e^-$ channel'). 
	Tables~\ref{table:eps_dxy} and \ref{table:eps_dxy_1} show, respectively for the $e^+e^- \ell^+\ell^-$ and $e^+e^-$ channels, the signal efficiencies and expected significances for various assumed values of $\varepsilon_e$, according to the $50~{\rm ab^{-1}}$ luminosity at Belle~II and the $e^+e^- \to X+J/\Psi$ cross section 
	\begin{eqnarray}
	\sigma(e^+e^- \to X + J/\Psi)=
	2.77\times 10^{-2} \times \left( \frac{\varepsilon_e}{10^{-3}} \right)^2 ~{\rm fb}\,.
	\end{eqnarray}
{ The significances are calculated from the expected $p$-value of background-only hypothesis for each case.}


	The final result of expected sensitivity with Belle~II at $50~{\rm ab^{-1}}$ luminosity is shown in Fig.~\ref{fig:sensitivity}.
	Also displayed in Fig.~\ref{fig:sensitivity} are the expected results with the $\eta_c$-related studies at Belle~II and BESIII as discussed in Sections III and IV.
	The displaced $e^+e^-$ vertex searches 
	can probe the 17~MeV $X$ boson in the region 
	$$
	2.5 \times 10^{-4}\leq \varepsilon_e \leq 8.0\times 10^{-4}
	$$ 
	with significance larger than 2 by assuming near-zero background,
	and it covers the $\varepsilon_e$ region preferred by Atomki.
	While we expect less than one signal event with the currently available Belle data sample of $1~{\rm ab^{-1}}$, we can start exploring the Atomki preferred region within a few years once Belle~II accumulates data sample of $10~{\rm ab^{-1}}$ or more.

	\begin{figure}[b!]
		\centering
		\includegraphics[height=3.2in]{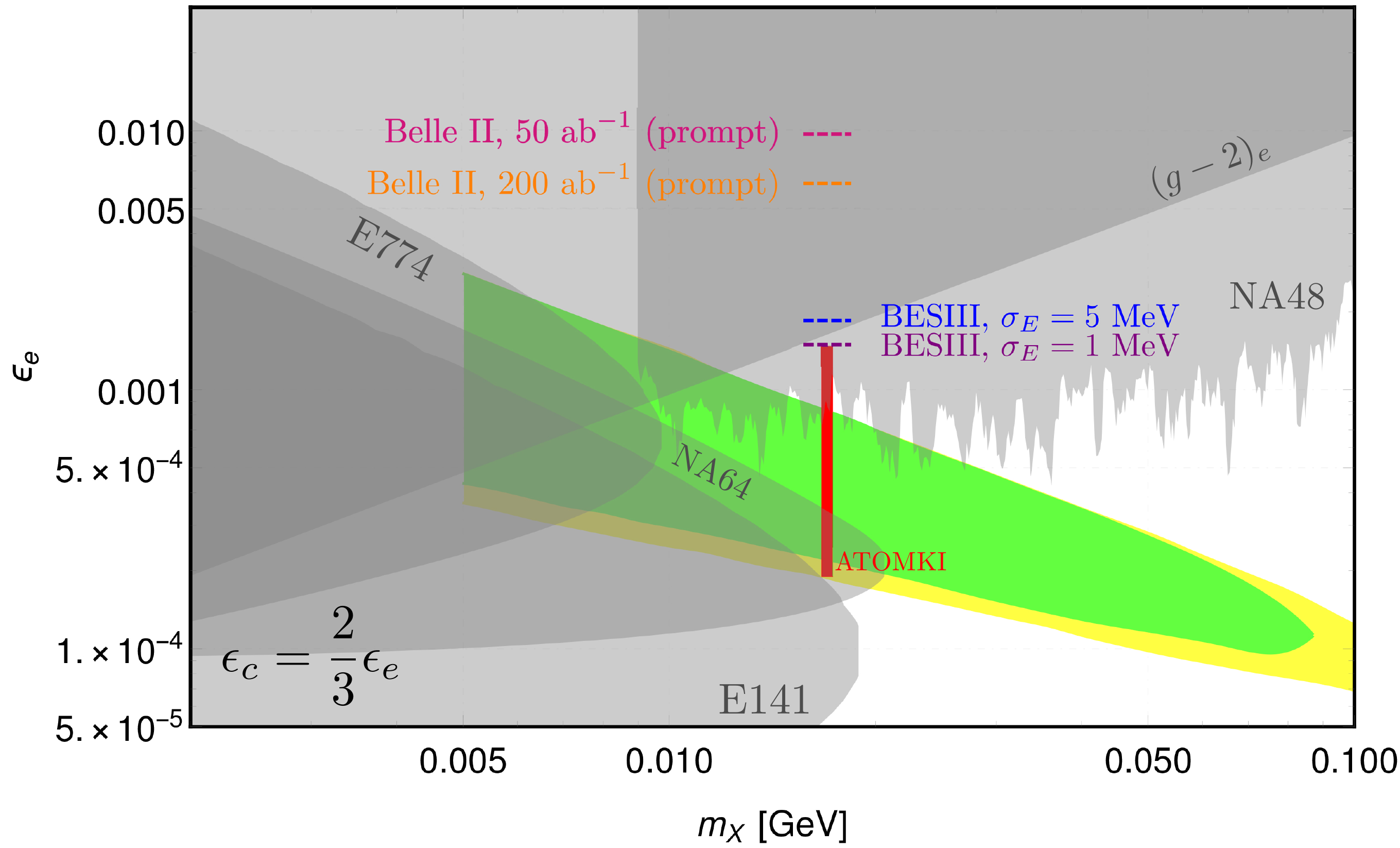}
		\caption{\small \label{fig:sensitivity}
			The {yellow (green)} contour corresponds 
			$\geq 2\sigma$ significance assuming SM background $B=0.1$ ($B=1$)
			from $e^+e^-$ channel at Belle~II
			with $50~{\rm ab^{-1}}$ luminosity,
			which probes rest of the favour parameter region of Atomki 
			(red vertical band).
			The gray-shaded regions are excluded by current experiments relate to $Xee$ coupling.
			Compare to the sensitivities of 
			$e^+e^- \to \ell^+\ell^- + J/\Psi \to \ell^+ \ell^- e^+ e^- \eta_c$ channel at Belle~II 
			and $e^+e^- \to J/\Psi \to e^+ e^- \eta_c$ channel at BESIII
			by assuming the kinematic mixing between $X$ boson and photon, 
			bringing the relation $\varepsilon_c=\frac{2}{3}\varepsilon_e$.
			For the future Belle~II experiments 50, 200 ab$^{-1}$ in the  17 MeV mass region, 
			the significance bounds are denoted by magenta, orange dotted lines, respectively.
			In the case of BESIII for the energy resolutions $\sigma_E = 1$ MeV and $5$ MeV,
			corresponding to section~\ref{sec:spectrum}, 
			they are denoted in purple and blue, respectively.
		}
	\end{figure}

	Our study with the displaced vertex is extended for wider mass range of $X$-like boson, whereby we determine the region of sensitivity with Belle~II at $50~{\rm ab}^{-1}$, as displayed in Fig.~\ref{fig:sensitivity}. 	
	Two cases are considered}: $B=0.1$ and $B=1$, where $B$ is the  expected number of background events from the SM processes}.  We then use the Poisson probability of the expected background to fluctuate upward, to calculate the $p$-values and the corresponding significances.
	The $2\sigma$ significance region with $B=0.1$ ($B=1$) is shown 
as the yellow (green) contour in Fig.~\ref{fig:sensitivity},
{which yields signal event $N_S \simeq 1\,(N_S\simeq 2)$}
\footnote{
In this case, the $p$-value is obtained as $\sum_{n \geq N_S+B} P(n|B)$ which is the probability of upward fluctuation based on the SM-only hypothesis with $P(n|B)$ being the Poisson probability of observing $n$ events given $B$ events of expected background. Then the confidence level (CL) is $\mathrm{CL} = 1 - p$. The $2\sigma$ significance corresponds to 95.4\% CL. For $B = 1~ (0.1)$ events, we need $N_S \simeq 2~ (1)$ signal events to attain $2\sigma$ significance.  
}.	
	This study can probe the parameter region of $5~{\rm MeV}\leq m_X \leq 100~{\rm MeV}$
	and $1.0\times 10^{-4}\leq \varepsilon_e \leq 3\times 10^{-3}$,
	which have not been constrained by any existing experiments.
	The upper edge of this region is determined by the condition
	$2~{\rm mm}\leq d_{xy}$, 
	while the lower edge is limited by the statistics. Therefore, with even higher luminosity of Belle~II exceeding the target $50~{\rm ab}^{-1}$, the lower edge of the sensitivity region can be extended further.

	\section{Summary and conclusion}
	\label{sec:summary}
	
	In summary, we propose several studies using $J/\Psi$ at lepton colliders such as Belle~II and BESIII, to search for light vector boson around the mass range suggested by the $^8{\rm Be}^*$ anomaly of the ATOMKI experiment.	At BESIII, the $J/\Psi \to \eta_c X \to \eta_c e^+e^-$ channel can be used 
	to constrain the vector boson and charm quark coupling, $\varepsilon_c$.  With the currently available sample of $N_{J/\Psi}=10^{10}$ and effective $\eta_c$ reconstruction efficiency of $1.23\%$, we can exclude the region $|\varepsilon_c| \gtrsim 5\times 10^{-3}$ for $m_X=17~{\rm MeV}$. 
	If $N_{J/\Psi}=10^{11}$ is produced at BESIII in the near future, exclusion of the region $|\varepsilon_c|\,\simeq\, 3\times 10^{-3}$ might be achieved.  If universal coupling to up-type quarks is assumed, we expect that	the entire favored signal region from the $^8{\rm Be}^*$ anomaly could be covered.
	
	On the other hand at Belle~II, with higher CM energy of 10.59~GeV, 
	we propose to study the process $e^+e^- \to \ell^+\ell^- J/\Psi$ followed by $J/\Psi \to \eta_c X \to \eta_c e^+e^-$.   Using the recoil mass against $\ell^+\ell^- e^+e^-$, 
	we perform MC study and find that the expected production of $J/\Psi$ events is about three orders of magnitude smaller than that of BESIII, thus yielding the sensitivity of $|\varepsilon_c|\, \gtrsim\, 1.8\times 10^{-2}$ at $m_X = 17~{\rm MeV}$.
	
	Alternatively, we can study the process $e^+e^- \to X + J/\Psi \to e^+e^- \ell^+\ell^-$ at Belle~II 
	and directly constrain  the $X$ boson-electron coupling.
	The $X$ boson is boosted by the higher CM energy and heavy mass of $J/\Psi$,
	producing displaced vertex of $X\to e^+e^-$ which is longer than several millimeters.
	Particularly, it is almost background free when the transverse flight distance is $2~{\rm mm}\leq d_{xy}\leq 8~{\rm mm}$.
	Selecting this window and requiring 
	$ > 2\sigma$ significance,
	it gives the sensitivity $2.0\times 10^{-4}\leq |\varepsilon_e| \leq 8.0 \times 10^{-4}$ at $m_X=17~{\rm MeV}$ for 50 ${\rm ab}^{-1}$ luminosity
	and covers most of the favored signal region from the claimed $^8{\rm Be}^*$ anomaly.
	Extending the range of the $X$ boson mass, this method can cover the unprecedented parameter space of $9~{\rm MeV}\leq m_X \leq 100~{\rm MeV}$ and $1.0\times 10^{-4}\leq |\varepsilon_e| \leq  10^{-3}$.

	\bigskip
	
	\section*{Acknowledgments}
	The work is supported in part by the National Research Foundation of Korea [NRF-2018R1A4A1025334 (SCP, YJK), NRF-2019R1A2C1089334 (SCP), and NRF-2020R1I1A1A01066413 (PYT)].
	%

	

	\bibliographystyle{JHEP}
	\bibliography{atomki_refs}

\end{document}